\documentclass[aps,floats,twocolumn]{revtex4-1}
\usepackage{amsmath}
\usepackage{bbm}
\usepackage{graphicx}
\usepackage[applemac]{inputenc}

\begin{document}

\title{Low-energy instability of flexural phonons in graphene}
\author{P. San-Jose$^a$, J. Gonz\'alez$^a$ and F. Guinea$^b$  \\}
\address{$^a$Instituto de Estructura de la Materia,
        Consejo Superior de Investigaciones Cient\'{\i}ficas, Serrano 123,
        28006 Madrid, Spain\\
        $^b$Instituto de Ciencia de Materiales de Madrid,
        Consejo Superior de Investigaciones Cient\'{\i}ficas, Cantoblanco,
        Madrid, Spain}

\date{\today}

\begin{abstract}
We study the effect exerted by the electrons on the flexural phonons in 
graphene, accounting for the attractive interaction created by the 
exchange of electron-hole excitations. Combining the self-consistent
computation of the phonon self-energy with renormalization group methods,
we show that graphene has two different phases corresponding to soft and
strong renormalization of the bending rigidity in the long-wavelength
limit. In the first case, the system may have an intermediate scale in
which the phonon dispersion is softened, but it manages finally to become
increasingly rigid over large distance scales. The strongly renormalized 
phase is closer however to critical behavior, with an effective rigidity
that becomes pinned in practice at very small values, implying a very
large susceptibility for the development of a condensate of the flexural
phonon field. 
\end{abstract}

\maketitle



In recent years, the feasibility of fabricating graphene 
layers a single atom thick\cite{novo,geim,kim}
has led to a great deal of activity, because of their novel electronic 
properties and potential applications\cite{rmp}.  
Beyond its remarkable electronic transport attributes, it is becoming increasingly clear that graphene's mechanical and electromechanical properties are just as remarkable. Despite it's extreme thinness, it behaves as a stable metallic membrane with finite elastic moduli and rigidity. From this point of view, one of the most intriguing features of graphene is its tendency to develop ripples with a characteristic length scale. In exfoliated graphene, these ripples are correlated to some extent with the irregularities of the 
substrate on which the graphene sheet is deposited. But there is also evidence
that they arise in part as an effect intrinsic to the two-dimensional membrane. 
When graphene was first fabricated, it came as a surprise that a purely 
two-dimensional material could exist, without being destabilized by large 
thermal or quantum fluctuations. From a theoretical point of view, 
it has been shown however that the coupling of flexural phonons to the 
in-plane vibrations in a crystalline membrane provides a mechanism to 
stabilize the two-dimensional system, leading to a finite critical temperature 
for the crumpling transition\cite{nelson}. The ripples are usually viewed then 
as a sort of imprint of the large fluctuations that should be present in the 
low-dimensional material\cite{rip}.

The ripples are expected to have a significant impact on the electronic 
transport in graphene\cite{kg}. More generally, it has been shown that the 
scattering with flexural phonons may lead to a reduction of the lifetime of 
electron quasiparticles in graphene\cite{mvo}. On the other hand, the 
effect exerted by the electrons on the out-of-plane vibrations of a 
metallic membrane has been much less studied. There have been partial 
approaches in which the coupling of the charge density to the flexural phonons 
has been considered\cite{ga1,njp,ga2}, but a zero temperature scaling analysis characterizing the 
possible critical behavior in the system has not been undertaken until now.

In this paper we investigate the many-body effects on flexural phonons in 
graphene, in a theory where these are coupled to the electron charge density as 
well as to the in-plane vibrations. These are known to give rise to a repulsive 
self-interaction between flexural phonons. We show that the exchange 
of electron-hole excitations leads to an interaction with the opposite character, although it plays in general the role of an irrelevant perturbation in the system. 
We will analyze the conditions in which this attraction may have a significant
effect, finding that it leads in that case to a softening of the acoustic branch 
of out-of-plane phonons in graphene. The effective bending rigidity can be then 
strongly suppressed beyond a certain length scale, making the system very 
susceptible to the development of a condensate (nonvanishing average value) for 
the out-of-plane phonon field.

We characterize the shape of the graphene sheet by the vector field 
${\bf u} = (u_1, u_2, h)$, where $u_1, u_2$ represent the in-plane displacement
with respect to the equilibrium position and $h$ is the out-of-plane shift. 
The deformation of the membrane is analyzed in terms of the strain tensor 
$u_{ij}$, given by 
\begin{equation}
u_{ij} = \frac{1}{2} (\partial_i u_j + \partial_j u_i + 
                       \partial_i h  \partial_j h  )
\end{equation}
Thus, the action of the system has a term governing the dispersion of
flexural phonons and depending on the mass density $\rho $ and the rigidity 
$\kappa $
\begin{equation}
S_{\rm free} = \frac{1}{2} \int dt \: d^2 x (\rho  (\partial_t h )^2 -
  \kappa (\nabla^2 h )^2  )
\label{s1}
\end{equation}
while the dynamics of the in-plane phonons is controlled by a term of the form
\begin{equation}
S_{\rm u} = \frac{1}{2} \int dt \: d^2 x (\rho  (\partial_t u_i )^2 -
      2\mu Tr \: u_{ij}^2 - \lambda  (Tr \: u_{ij})^2  )
\label{s2}
\end{equation}
where $\mu $ and $\lambda $ are the Lam\'e coefficients.     

The term $S_{\rm u}$ contains a quartic self-interaction for the out-of-plane 
phonon field $h$. It is well-known that this interaction has to be taken into 
account for the description of stable crystalline membranes, as the soft 
dispersion in (\ref{s1}) leads to infrared divergences that cannot be tamed 
in the free theory\cite{nelson}. 
The novel effect arising in graphene is that the electronic 
charge couples to the displacement of the sheet, in such a way that it
gives rise to another source of self-interaction between the phonon fields.
We will consider the effect of electrons from the $\pi $ bands of graphene,
represented by the field $\Psi (x)$. Accounting for the fact that the coupling
must take place through the trace of the strain tensor $u_{ij}$, we add
to the action a new term
\begin{equation}
S_{\rm e-ph} = -g\int dt \: d^2 x \; \Psi^{\dagger} (x) \Psi (x) \: Tr \: u_{ij}
\label{s3}
\end{equation}
The electron-phonon coupling $g$ has dimensions of energy and is directly
related to the deformation potential of the graphene lattice, which has a microscopic origin.

Assuming that the electron correlations are weak in graphene, we may represent 
the dynamics of the $\Psi (x)$ field by a quadratic hamiltonian, matching at 
low energies the usual expression for the Dirac fermion fields. The electron 
degrees of freedom can be integrated out at this point, leading from (\ref{s3}) 
to a new phonon contribution to the action, to be added to (\ref{s1}) and 
(\ref{s2}),
\begin{equation}
S_{\rm u'} = -\frac{1}{2} g^2 \int d\omega_q \: d^2 q  \; \chi ({\bf q},\omega_q )
        (Tr \: u_{ij})  (Tr \: u_{ij})
\label{u2}
\end{equation}
where $\chi ({\bf q}, \omega_q )$ is the density-density correlator depending on 
frequency and momentum variables. Terms that are higher order in $Tr \: u_{ij}$ 
can be generated after integration of the electron fields, but they turn out 
to be more irrelevant perturbations in the approach developed in what follows.

Since the energy scale of electronic excitations (for a given momentum) are much larger than the phonon energies, we can safely take the limit $\omega = 0$ in the charge 
susceptibility. In the Dirac theory, we have in particular 
$\chi ({\bf q}, 0 ) = -|{\bf q}|/v_F$. We observe that the new contribution
(\ref{u2}) tends to cancel the term with the LamÈ coefficient $\lambda $ in (\ref{s2}). 
After following the usual procedure to integrate the in-plane phonons 
$u_i$ \cite{nelson}, we get the total action for the flexural phonons
\begin{eqnarray}
S  & = &   \frac{1}{2} \int d\omega_q \: d^2 q (\rho \: \omega_q^2 h^2 -
                                 \kappa {\bf q}^4 h^2  )        \nonumber      \\
  &  &  - \frac{1}{2} \int d\omega_q \: d^2 q \; 
  K(q) \widetilde{u}_{ij}({\bf q}, \omega_q) \widetilde{u}_{ij}({-\bf q}, -\omega_q)
\label{act}
\end{eqnarray}
where $\widetilde{u}_{ij}({\bf q}, \omega_q)$ is the Fourier transform of 
$(1/2) P_{ij} \partial_i h  \partial_j h$, $P_{ij}$ being the transverse 
projector, and the quartic coupling is (in the Dirac theory) 
\begin{equation}
K(q) = \frac{4\mu(\mu+\lambda - g^2 |{\bf q}|/v_F)}{2\mu + \lambda - g^2 |{\bf q}|/v_F }
\label{kq}
\end{equation}

Negative values of the coupling function $K(q)$ may lead to an instability of 
the graphene sheet, as they represent an attractive interaction between the 
flexural phonons. The phonon self-energy and the interaction vertices are 
actually affected by logarithmic divergences in the energy cutoff, which makes 
it necessary to keep track of the low-energy dependence of the couplings. In order 
to study the competition between positive and negative couplings in (\ref{kq}),
we may approximate this coupling function by the constant term 
$K_0 = 4\mu (\mu + \lambda )/(2\mu + \lambda )$ and the dominant powers of 
$|{\bf q}|$
\begin{equation}
K(q) \approx K_0 - G_1 \: |{\bf q}| - G_2 \: {\bf q}^2
\label{linear}
\end{equation}
Neglecting higher-order terms in $|{\bf q}|$ amounts to disregarding the exchange 
of an increasing number of electron-hole pairs in the in-plane phonon propagator, which is justified 
as long as the pole arising from the denominator of (\ref{kq}) is not included within the first Brillouin zone of a real 
graphene system. This is consistent with an estimate for the on-site deformation 
potential for the in-plane phonons $g_{\rm in} \approx $ 8 eV \cite{jiang}, 
that is the coupling appearing in the 
denominator of Eq. (\ref{kq}). 

\begin{figure}[t]
\begin{center}
\includegraphics[width=3.2cm]{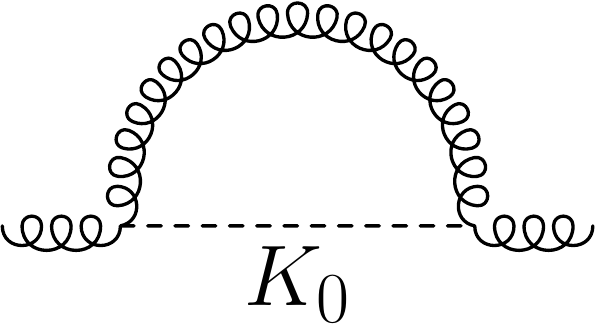}
\hspace{0.8cm}
\includegraphics[width=3.2cm]{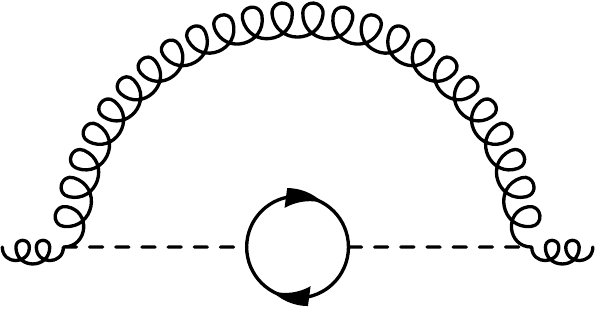}\\\vspace{0.5cm}
 \hspace{0.3cm}  (a) \hspace{3.6cm} (b) 
\end{center}
\caption{First order corrections to the self-energy of flexural phonons 
(represented by a wiggly line) arising from (a) the four-phonon interaction
$K_0$ and (b) the exchange of electron-hole excitations (represented by the 
bubble with arrow lines).}
\label{one}
\end{figure}

The different interactions tend to modify the low-energy behavior
of the rigidity $\kappa $ in opposite directions depending on their attractive
or repulsive character. This can be seen through the corrections to the phonon
self-energy $\Sigma ({\bf p}, \omega_p )$ represented in Fig. \ref{one}.
Computing the full propagator of the flexural phonons from the expression 
$D^{-1} ({\bf p}, \omega_p ) = \omega_p^2 - (\kappa_0 /\rho) {\bf p}^4 
 - \Sigma ({\bf p}, \omega_p )$, we find the renormalized bending rigidity
$\kappa ({\bf p})$ given by the equation
\begin{eqnarray}
\lefteqn {     \frac{\kappa ({\bf p})}{\rho }  =  
  \frac{\kappa_0 }{\rho }  + \frac{1}{8 \pi^2 }
 \int_0^{q_c } dq \int_0^{2\pi } d\phi  }                         \\
   &  &   \;\;\;\;\;\;\;\;    \sin^4 (\phi)
   \frac{K_0 - G_1 |{\bf p}-{\bf q}| - G_2 |{\bf p}-{\bf q}|^2}{\rho^2 |{\bf p}-{\bf q}|^4}
   \frac{|{\bf q}|^3}{\sqrt{\kappa({\bf r}) /\rho  }} \nonumber
\label{integ}
\end{eqnarray}
A self-consistent treatment of the many-body theory requires that the rigidity 
arising in the integrand of (\ref{integ}) from the phonon propagator must be also
taken as $\kappa ({\bf p})$. This leads to an integral equation, that
can be solved to obtain the effective momentum dependence of the renormalized 
parameter.

\begin{figure}[t]
\begin{center}
\includegraphics[width=4.2cm]{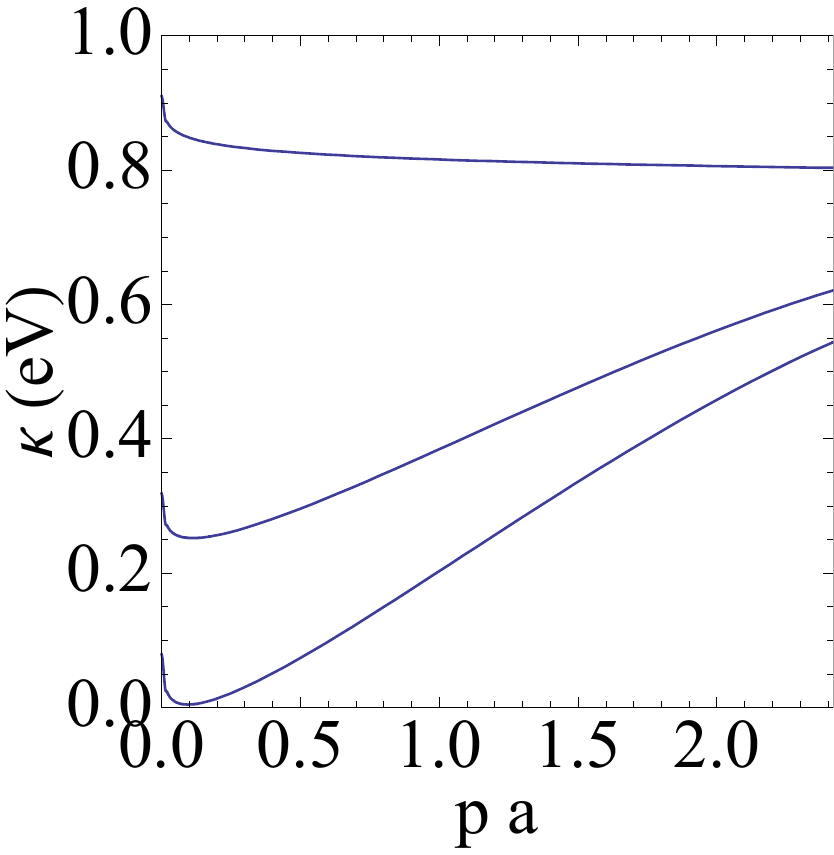}
\includegraphics[width=4.2cm]{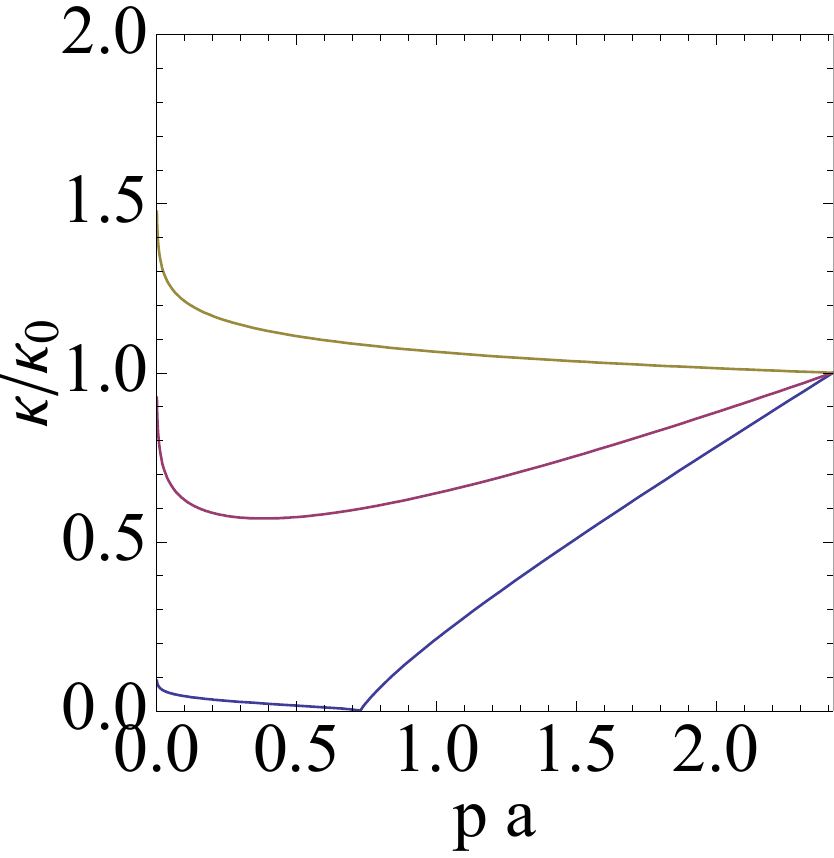}\\
 \hspace{0.3cm}  (a) \hspace{3.6cm} (b) 
\end{center}
\caption{(a) Effective momentum dependence of the bending rigidity obtained
from the solution of Eq. (\ref{integ}) for values of $G_1$ and $G_2$ corresponding 
(from top to bottom) to a deformation potential $g$ of flexural phonons equal 
to 0, 24 and 26.3 eV. 
(b) Scaling
of $\kappa $ in the renormalization group approach, for bare values of the 
deformation potential equal (from top to bottom) to 0, 22 and 29 eV. In both 
approaches, we have taken the bare value of $a^2 K_0$ equal to 18 eV.}
\label{two}
\end{figure}

We have represented in Fig. \ref{two}(a) the result of solving Eq. (\ref{integ})
in different regimes of the coupling $g$. We observe that, for a 
sufficiently large value $g_c$, the function $\kappa ({\bf p})$ vanishes at 
a certain momentum, bouncing back for smaller values of $|{\bf p}|$. In this 
picture, $g_c$ plays the role of critical coupling, as larger values of 
$g$ would actually lead to negative values of the bending rigidity, 
implying the absence of a self-consistent solution. In that case, however,
a more refined analysis has to be applied in order to capture the low-energy 
scaling of the couplings $K_0$, $G_1$ and $G_2$, which becomes quite significant
when the effective bending rigidity gets very small.

We apply then a Wilsonian renormalization group method starting from a 
high-energy cutoff $E_c = \sqrt{\kappa_{0} /\rho } \: q_c^2$ for some large 
momentum $q_c$, and performing a progressive integration of energy shells 
to approach the low-energy regime. The bare coupling $K_0$ 
is corrected to lowest order by the one-loop diagram made of the exchange of 
two flexural phonons, which 
shows a logarithmic dependence on the cutoff $q_c$. This can be absorbed into a
suitable renormalization of the effective coupling. Under a differential change 
of the energy scale, we find the scaling with the running cutoff 
$q \rightarrow 0$ 
\begin{equation}
q \frac{\partial (K_0/\rho^2) }{\partial q } = 
  \frac{3}{64 \pi} \frac{(K_0/\rho^2)^2}{(\kappa/\rho)^{3/2} }
\label{k0rg}
\end{equation}

The couplings $G_1$ and $G_2$ correspond to effective phonon interactions 
mediated by the exchange of electron-hole excitations, and they also undergo a 
logarithmic renormalization from the exchange of flexural phonons.
However, the main source of scaling for these couplings arises at the classical
level, as the respective terms in the effective action  
pick up anomalous factors $1/\sqrt{s}$ and $1/s$ when shrinking the energy scale from 
$E_c$ to $E_c/s$. We find then the respective anomalous dimensions $\Delta = -1/2$ and 
$-1$, which lead to the scaling equations for $n = 1, 2$
\begin{equation}
q \frac{\partial G_n }{\partial q } =  n \: G_n  
  + \frac{3}{32 \pi} \frac{(K_0 /\rho^2 ) G_n}{(\kappa/\rho)^{3/2} }
\label{grg}
\end{equation}  
The main scaling dependence is given by the behavior
\begin{equation}
G_n (q )  = q^n \: \widetilde{G}_n (q )
\end{equation}
where
the couplings $\widetilde{G}_n$ are subject to a purely logarithmic 
scaling in the same fashion as $K_0$.

A scaling equation can be also written for the rigidity $\kappa $, as the 
self-energy diagrams in Fig. \ref{one} actually display a logarithmic dependence
on the cutoff $q_c$. Taking into account the irrelevant character of the
couplings $G_n$, we find that under a variation of the momentum scale $q$
\begin{equation}
q \frac{\partial \kappa /\rho }{\partial q } = 
 - \frac{3}{16 \pi} \frac{K_0/\rho^2}{(\kappa/\rho)^{1/2} }
 + \frac{3}{16 \pi} \sum_n \: q^n \: \frac{\widetilde{G}_n/\rho^2}{(\kappa/\rho)^{1/2} }
\label{kprg}
\end{equation}
Eq. (\ref{kprg}) has to be solved then together with (\ref{k0rg}) and 
(\ref{grg}) in order to obtain the behavior of the effective rigidity 
$\kappa $. We note that, in real graphene samples, the bare values of the 
couplings $K_0$ and $G_n$ place in general the system in the weak-coupling 
regime. On the one hand, the energy scale of $K_0$ is given by  
$a^2 4\mu(\mu + \lambda )/(2\mu+\lambda)\approx 20$ eV \cite{fas}, where $a=1.4~\textrm{\AA}$ is the interatomic spacing. The 
coupling from the electronic sector is subject to the uncertainty in 
the estimates of the deformation potential for out-of-plane vibrations, leading 
to a wide range of values with $g \sim 20-30$ eV \cite{ando}. On the other hand,
we have the energy scale $a^2 \rho^2 (\kappa /\rho )^{3/2}$, which has a nominal 
value of $\approx 40$ eV in graphene. When $\kappa/\rho$ becomes very small, 
however, the rate of decrease of $K_0, \widetilde{G}_1$ and $\widetilde{G}_2$ 
is greatly increased, leading to significant differences with respect to the 
behavior of the effective rigidity obtained from the self-consistent equation 
(\ref{integ}).

The behavior of $\kappa $ obtained from the scaling equations is shown in 
Fig. \ref{two}(b) for different bare values of $g$. For small values of this 
coupling, we see that the effective rigidity has a steadily increasing trend 
as $q \rightarrow 0$. There is however a regime of large $g$, 
corresponding to values of the deformation potential $g \sim 25-30$ eV, in 
which $\kappa $ is significantly reduced due to the electron-phonon 
interaction. At some point, $\kappa $ may be even suppressed by
several orders of magnitude, but without vanishing as in the 
self-consistent approach shown above. Here we find instead that $\kappa $
always bounces back from zero, though it may become trapped in a range of very small
values as the effective couplings $K_0, G_1$ and $G_2$ are themselves quickly 
renormalized very close to zero.

We find therefore that graphene has two different phases corresponding 
to soft and strong renormalization of $\kappa $ as $q \rightarrow 0$. In the 
first case, the system may have an intermediate scale in which the effective 
$\kappa $ is reduced, but it manages finally to become increasingly rigid over 
large distance scales. In the strongly renormalized phase, however, the
effective rigidity becomes pinned in practice at very small values below
a certain momentum scale. 

\begin{figure}[t]
\begin{center}
\includegraphics{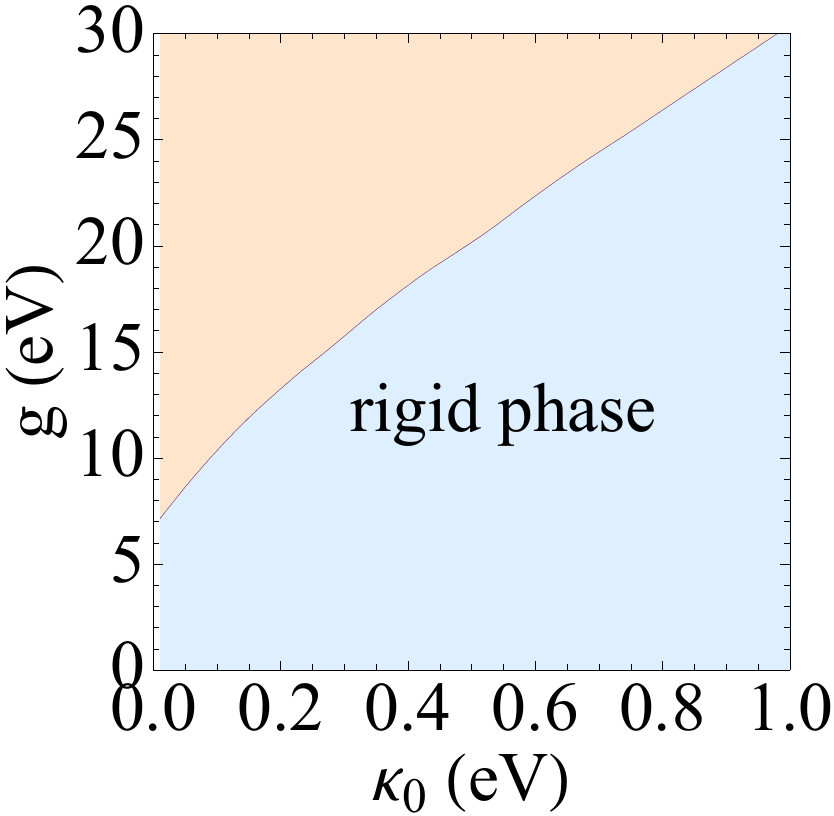}
\end{center}
\caption{Diagram showing the phases corresponding to a rigid long-distance 
regime of graphene (lower region) and pinning of the renormalized bending
rigidity at very small values (upper region), in the space of nominal values 
of the rigidity $\kappa_0$ and deformation potential $g$. The value of $a^2K_{0}$ was taken as $18$ eV.}
\label{three}
\end{figure}

Although the self-consistent solutions of Eq. (\ref{integ}) miss the 
renormalization of the couplings at very small $\kappa $, we can still use 
them to draw the boundary between the phases with soft and strong 
renormalization of the bending rigidity. That line can be formed with the
values of critical coupling $g_c$ beyond which the integral equation
would lead to negative values of $\kappa $. This forbidden region
corresponds in the renormalization group approach to the regime where the
bending rigidity becomes trapped at very small values. The resulting
phase diagram in the space of nominal values of the rigidity $\kappa_0$
and deformation potential $g$ is shown in Fig. \ref{three}.

The phase with strong renormalization of the bending rigidity can be put 
in correspondence with the appearance of ripples in graphene, as very small
values of $\kappa $ imply a very large susceptibility for the development of 
a condensate (nonvanishing average value) of the flexural 
phonon field $h$. It is conceivable that the different degree of 
Coulomb screening may control which of the two phases is found in a real
graphene sample, as the Coulomb interaction leads to a slight reduction  
in the charge susceptibility. Thus, graphene can be more prone to develop 
ripples or to buckle when surrounded by media with high dielectric constant, 
for which the values of $\chi ({\bf q}, 0 )$ are maximized.

\end{document}